\documentclass[prl,aps,showpacs,twocolumn,floatfix]{revtex4}
\usepackage{epsf}
\usepackage[dvips]{color}
\usepackage{graphicx}

\begin{document}

\title{Dielectric microscopy of with submillimeter resolution}

\author{Nathan S. Greeney and John A. Scales}
\affiliation{Department of Physics, Colorado School of Mines, Golden
Colorado 80401}

\begin{abstract}
In analogy with optical near-field scanning methods, we use tapered dielectric 
waveguides as probes for a millimeter wave vector network analyzer.  By 
scanning thin samples between two such probes we are able
to map the spatially varying dielectric properties of materials with 
sub-wavelength resolution; using a 150 GHz 
probe in transmision mode we see spatial resolution 
of around 500 microns. 
We have applied this method to a variety of highly heterogeneous materials.
Here we show dielectric maps of granite and oil shale.
\end{abstract}

\maketitle

\section{Introduction}

Near-field scanning optical microscopy (NSOM) is an 
established technique for
achieving nano-scale resolution of surface material properties
\citep{nsombook}.  
Recently these ideas have been extended to millimeter 
waves;
that is, electromagnetic waves spanning roughly from 30 to 300
GHz (e.g., \citep{nozokido:148}, \citep{kume:056105}).
In particular,
Kume and Sakai
\citep{kume:056105} used a dielectric probe machined from Teflon.
We apply a very similar technique with conical Teflon probes on either
side of thin sections of rocks.  This allows us to achieve transmission-mode 
dielectric maps of rocks with submillimeter resolution, limited in speed only
by the mechanical scanning stages.

\section{Setup}

\begin{figure}
\centerline{
\includegraphics[width=65mm]{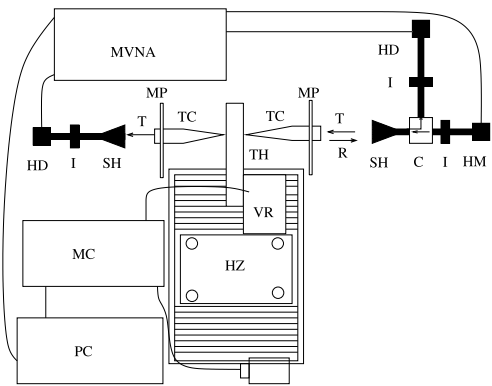}}
\caption{\footnotesize\textit{Experimental Setup.} Microwaves (8-18 GHz)
  are generated by the Vector Network Analyzer (VNA) and converted into
  millimeter waves via a harmonic multiplier (HM).
  The Teflon cones (TC) guide the millimeter waves to a point as they
  pass through the sample in the Teflon holder (TH). The harmonic
  detectors (HD) receive the transmitted (T) and reflected (R)
  E field.  The VNA is controlled by a computer via LabView. 
The same computer also controls the x-z scanning stage.
  Other components in
  the setup are metal plates (MP), a circulator (C), isolators
  (I), and scalar (or corrugated) horns (CH).}
\label{figure1}
\end{figure}

Fig.\,\ref{figure1} shows the experimental setup, which is 
based on a
millimeter wave vector network analyzer (MVNA) made by AB Millimetre
of France.
The MVNA generates stabilized centimeter waves (8-18 GHz) phase
locked to a rubidium standard
which are then converted to millimeter waves
(MMW) as they pass through a harmonic multiplier. From here
the MMW propagate via single mode wave guide and are radiated
from a scalar (or corrugated) horn.  In the wave guide the E field 
is in a TE01 mode and is vertically polarized.  The effect of the 
scalar horn is to produce a axially symmetric radiation pattern with
no cross-polarization.  This is a hybrid mode \citep{goldsmith}.

The free-space MMW are then coupled into a Teflon cone which
acts as a near-field probe placed near the surface of the sample.
The cone is inserted through a small opening in an aluminum plate
to prevent
diffraction around the probe.
Teflon is used due to its
high transparency at microwave and millimeter wave frequencies. 
The E field transmitted through the sample is coupled to another
Teflon cone, which in turn couples to a scalar horn.  The transmitted
field is then
detected via a harmonic detector driven by a local oscillator.
The
reflected wave goes back through the first cone and scalar horn
and then
via a circulator to a second harmonic detector.  All the microwave 
oscillators are phase locked to the Rubidium clock.  For more details 
see \citep{scales_batzle_apl2}, \citep{mmwaves_expt}.

Holding the sample between the cones is a Teflon holder attached to
two linear motors. One is positioned vertically while the other is
positioned horizontally. Both of the motors are connected 
to a Newport controller, which is 
controlled by a computer running
{\it LabView}.

The idea behind the measurement is that as the probe scans the surface
of a heterogeneous sample, the phase of the transmitted E field is
perturbed in proportion to changes in the local index of refraction
(assuming a constant thickness sample).
Hence a spatial map of the phase of the E field can be directly translated
into a spatial map of local variations in the index of refraction.
Perturbations in the
index of refraction 
(a size-independent material property) can be related to perturbations
in the transmitted phase via
$$
\delta \phi = \frac{2 \pi d}{\lambda} \delta n
$$
where $d$ is the thickness of the sample.
In all cases the measured phase angles have been unwrapped.

\section{Experiment}

\subsection{Benchmark:  perfboard}

\begin{figure}
\centerline{
\includegraphics[width=75mm]{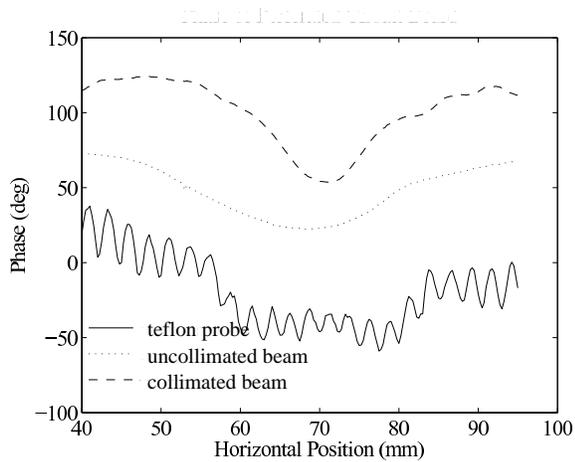}}
\caption{\footnotesize\textit{Phase contrast of perforated 
(synthetic-resin-bonded paper (SRBP) circuit board.} 
Shown here is the unwrapped phase of the 150GHz E field transmitted
   through the perfboard with 6 layers of masking 
  tape covering the holes from about 58mm to 82mm. The perfboard
  has holes
  drilled into it in a 2.56mm grid spacing. The solid line shows the phase
measured with  two Teflon cones as wave guides, 
the doted line shows phase measurements using polyethylene collimating
lenses, and the dashed line is using an uncollimated beam.
  The Teflon cones increase the resolution to the point of being able
  to resolve the holes in the circuit board.}
\label{figure2}
\end{figure}

As a first test of the system we use a piece of perforated 
circuit board (a synthetic resin bonded paper with no 
conductive soldering tabs) as a benchmark.  The soldering
holes
(Fig.\,\ref{figure2}) are drilled on a 2.56 mm centers.
In addition to the holes, we put 6 layers of masking tape 
on the perfboard to create a linear feature.

As can be seen from Fig.\,\ref{figure2} after scanning across
the sample, the holes themselves
cannot be seen with either the collimated or uncollimated
150 GHz beam.  The index change due to the tape is smeared out
over a distance of around 20 mm, which is the beam width.
However, with the Teflon probes both the holes and the tape
can be easily resolved.  The free-space wavelength
at 150 GHz is 2 mm.  Fig.\,\ref{figure2} suggests a resolution
of at least $\lambda /4$, or around 500 $\mu$m.

\begin{figure}
\centerline{
\includegraphics[width=75mm]{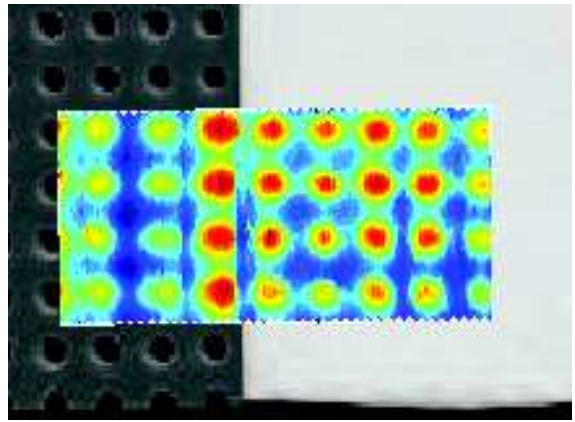}}
\caption{\footnotesize\textit{Spatial Phase Map of Perforated Circuit
    Board.} Shown here is the phase change spatially of a 150GHz
  transmitted wave through a piece of perforated circuit board with 6
  layers of tape covering the holes as shown above.  The 
  Teflon probes were used.
  The phase map is overlaid on an optical scan
  of the perfboard.}
\label{figure3}
\end{figure}

With the ability to
see the holes in a linear scan, we moved up to a two dimensional scan
to produce Fig.\,\ref{figure3}, which shows an overlay of the
phase contrast map and an optical scan of the perfboard.
In the two dimensional scan, you can
clearly see the holes and also the change from only circuit board to
circuit board with tape.

\subsection{A strongly heterogeneous medium}

\begin{figure}
\centerline{
\includegraphics[width=80mm]{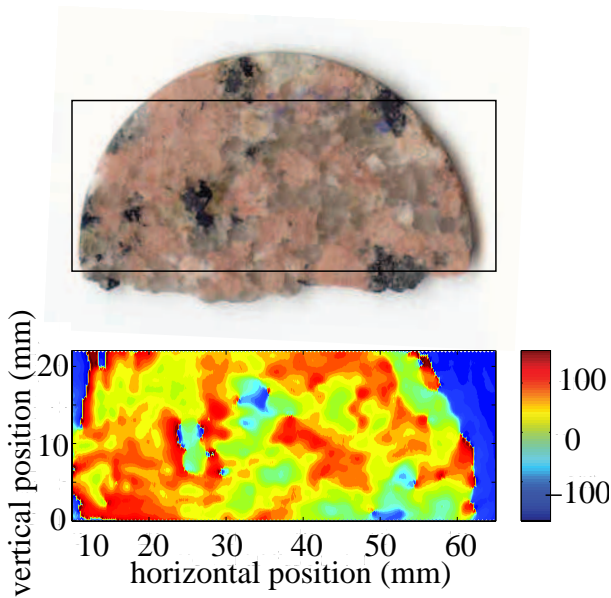}}
\caption{\footnotesize\textit{Phase Contrast Map of Granite.} Shown
  here is the phase of a 150GHz E field transmitted through a 2.13 mm thick
  piece of granite (below) and an optical scan of the granite sample
  (above). Changes in the phase relate to changes in index of
  refraction therefore allowing us to determine local changes in the
  dielectric properties of granite. 
  The red areas correspond to quartz grains, the green areas to feldspar and
  the blue areas to biotite.}
\label{figure4}
\end{figure}

After these preliminary benchmarks we next
scanned some granular materials that we have previously used in studies
of wave propagation in random media (\citep{scales_batzle_apl2}, 
\citep{mmwaves_expt}, \citep{PhysRevE.67.046618}, 
\citep{malcolm:015601})
For the first scan we
chose to use a piece of medium grained granite (Llano granite).
In the top of Fig.\,\ref{figure4} an optical scan of the sample
shows the main constituents of this granular sample: the black grains are
biotite (a common mineral within the mica group), 
the gray grains are quartz (silica)
and the pink grains are feldspar (a common mineral
crystallized from magma that makes up a large fraction of the Earth's crust).

Scanning over a piece that is uniformly 2.13 mm thick gave us
Fig.\,\ref{figure4}.  Correlating the optical scan and the phase
map it is apparent that the red colors (large positive phase
perturbations) corresponds to quartz grains.  The green color (relative 
phase of zero degrees) corresponds to Feldspar.  The bluish color (large
negative phase perturbation) corresponds to biotite.  Since the sample is
about 2 mm thick (i.e., one $\lambda$), a $\pi/2$ phase change corresponds
to an index change of about $\frac{\pi/2}{ 2 \pi} = .25.$
Previously measured 
\citep{vaccaneo} microwave values of the index for Quartz (2.05 - 2.23)
and Mica (2.16 - 3.0) are consistent with these measured phase changes.

\subsection{A finely laminated medium}

\begin{figure}
\centerline{
\includegraphics[width=80mm]{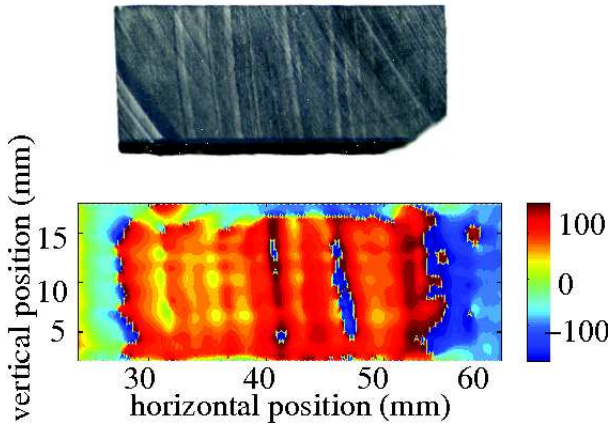}}
\caption{\footnotesize\textit{Phase Contrast Map of Oil Shale.} Shown
  here is the phase of a 150GHz E field transmitted through a rectangular
  piece of oil shale (below) and an optical scan of the oil shale
  sample (above). The diagonal markings on the optical scan are kerf
marks from the cutting tool.  They don't show up in the phase map.}
\label{figure5}
\end{figure}

A  sedimentary rock, such as oil shale, can show
millimeter scale lamination.  The birefringence caused by
this layering was previously discussed in \citep{scales_batzle_apl2}.
By doing the same scan on a 4.44 mm thick piece of oil shale as we did
on the granite, we were able to see changes in the phase when moving
from organic-rich areas to organic-poor areas in the oil shale. As
Fig.\,\ref{figure5} shows, we were able to resolve dielectric changes
of less than a millimeter across.  The phase change in this figure going
from blue to red is about 3.5 radians.  This corresponds to an index change
of $3.5 * 2 /(2 \pi 4.44) \approx .25$  This is very close to the index
differences seen \citep{scales_batzle_apl2} for homogeneous 
samples of oil-rich and oil-poor shale.

\section{Conclusion}

By using an easily manufactured conical Teflon probe we were able
to greatly enhance the spatial resolution of our quasi-optical 
millimeter wave setup via a near-field scanning probe technique.
This continuous wave
technique has extremely high sensitivity to small index-induced
phase changes of the transmitted electric field.
We tested this technique on strongly heterogeneous
granular samples (metamorphic and sedimentary rocks) and were
able to obtain resolution of around 500 microns using a 150 GHz probe.
The technique
is fast (limited only by the speed of the mechanical scanning) and is
easily applied to any free space quasi-optical system.  Many other applications
suggest themselves, including near-field measurements of planar antennae,
integrated circuits, and other amorphous materials.

\textbf{Acknowledgements}

This material is based upon work supported by the National Science
Foundation under Grant EAR-0337379.

\bibliographystyle{prsty}
\bibliography{main}

\end{document}